\documentclass[]{aa}
\usepackage{graphicx}
\newcommand{\be}{\begin{equation}}
\newcommand{\ee}{\end{equation}}
\begin{document}

\title{Variable gamma-ray emission from the Be/X-ray transient A0535+26?}

\author{Gustavo E. Romero\thanks{Member of CONICET}, M.M. Kaufman Bernad\'o, Jorge A. Combi$^{\star}$, Diego F. Torres}

\offprints{G.E. Romero}

\institute{Instituto Argentino de Radioastronom\'{\i}a, C.C.5,
(1894) Villa Elisa, Buenos Aires, Argentina}

\date{Received / Accepted}
\abstract{We present a study of the unidentified gamma-ray source
3EG J0542+2610. This source is spatially superposed to the
supernova remnant G180.0-1.7, but its time variability makes
unlikely a physical link. We have searched into the EGRET location
error box for compact radio sources that could be the low energy
counterpart of the gamma-ray source. Although 29 point-like radio
sources were detected and measured, none of them is strong enough
as to be considered the counterpart of a background gamma-ray
emitting AGN. We suggest that the only object within the 95\%
error box capable of producing the required gamma-ray flux is the
X-ray transient A0535+26. We show that this Be/accreting pulsar
can produce variable hadronic gamma-ray emission through the
mechanism originally proposed by Cheng \& Ruderman (1989), where a
proton beam accelerated in a magnetospheric electrostatic gap
impacts the transient accretion disk. }

\titlerunning{Gamma-rays from A0535+26}
\authorrunning{G.E. Romero et al.}

\maketitle

\section{Introduction}
The EGRET instrument of the late Compton Gamma Ray Observatory
detected 271 point-like gamma-ray sources at energies $E>100$ MeV
during its lifetime (Hartman et al. 1999). The majority of these
sources still remain unidentified. Excluding 6 possible artifacts,
there are 75 sources at low galactic latitudes $|b|<10^{\circ}$
still not conclusively identified with objects seen at lower
energies (Romero et al. 1999). Several of these low-latitude,
presumably galactic sources seem to display significant levels of
variability over timescales of months (Tompkins 1999, Torres et
al. 2001a, b). A few particular cases of variable gamma-ray
sources at low latitudes have been recently studied by Tavani et
al. (1997, 1998), Paredes et al. (2000) and Punsly et al. (2000),
but there is not yet consensus on their nature. The variety of
behaviours displayed by the sources seems to suggest that there
exist more than a single class of gamma-ray emitting objects in
the Galaxy. To establish the nature of these peculiar objects is
one of the most urgent problems of high-energy astrophysics.

In this paper we present a study of the gamma-ray source 3EG
J0542+2610. We shall show that the only known object within the
95\% confidence location contour of the source capable of
generating the observed gamma-ray emission is the Be/X-ray
transient A0535+26. We shall argue that the gamma-rays are
produced during the accretion disk formation and loss phases in
each orbit, through hadronic interactions between relativistic
protons accelerated in an electrostatic gap in the pulsar
magnetosphere and the matter in the disk. The structure of the
paper is as follows. In the next section we summarize the main
characteristics of 3EG J0542+2610 and the results of our analysis
of the radio field within the EGRET location error contours. We
then present the peculiarities of the system A0535+26. In Section
4 we present a model for the gamma-ray production based on the
accelerator gap model by Cheng \& Ruderman (1989, 1991). The main
difference between our treatment and the accreting pulsar
gamma-ray model of Cheng et al. (1991) lies in the introduction of
time variability through the variation of the matter content of
the disk, which is a transient structure according to recent
observations.

\section{The gamma-ray source 3EG J0542+2610 and the surrounding radio field}

The best estimated position of the gamma-ray source 3EG J0542+2610
is at ($l$, $b$)$\approx$(182.02,$-1.99$). Its 95\% confidence
location contour overlaps with the shell-type supernova remnant
(SNR) G180.0-1.7 (Romero et al. 1999, Torres et al. 2001c). The
gamma-ray flux for the combined EGRET viewing periods is
$(14.7\pm3.2)\;10^{-8}$ ph cm$^{-2}$ s$^{-1}$. Since the flux is
highly variable on timescales of months, it should be originated
in a compact object and not in the extended SNR. In Figure 1 we
show the EGRET light curve. The source switches between periods of
clear detections and periods when only upper bounds to the flux
can be determined. The variability analysis of Torres et al.
(2001b) assigns to 3EG J0542+2610 a variability index $I=3.16$,
which means that the source variability level is $4.32\sigma$
above the average (spurious) variability of all known gamma-ray
pulsars. Tompkins' (1999) variability index for this source
($\tau=0.7$) also indicates that the source is variable.

If a pulsar origin is discarded due to the high variability and
the steep spectral index ($\Gamma=-2.67\pm0.22$), we are left with
two main possibilities: 1) the source is a background, unnoticed
gamma-ray blazar seen through the galactic plane, or 2) it is a
galactic compact object with an energy budget high enough as to
generate significant gamma-ray emission and, at the same time, has
not the stable properties usually associated with isolated
pulsars.

In the standard model for gamma-ray blazars, the high energy
photons are produced by ultra-relativistic electrons through
inverse Compton interactions with seed lower energy photons that
can come from external sources (e.g the accretion disk) or from
the jet itself (see Krolik 1999 and references therein). The model
is supported by the fact that all identified gamma-ray AGNs are
strong flat-spectrum radio sources (Mattox et al. 1997).

\begin{table}
\caption[]{Point radio sources within the inner location
probability contours of the gamma-ray source 3EG J0542+2610}
\begin{tabular}{ccccc}
\hline
\\
Number & $l$ & $b$ & $S_{1.42\;{\rm GHz}}$ & Spectral\\ (on the
map) & (deg) & (deg) & (mJy) & index\\
\\
\hline
\\
1&181.46&-1.56&18.00&\\
2&181.50&-2.18&13.15&$\alpha_{1.42}^{0.408}=-2.9$\\
3&181.56&-1.71&62.90&$\alpha_{1.42}^{0.408}=-1.3$\\
4&181.59&-1.75&78.67&$\alpha_{1.42}^{0.365}=-1.6$\\
&&&&$\alpha_{4.85}^{1.42}=-0.9$\\ 5&181.61&-2.22&7.65&\\
6&181.62&-2.09&7.39&\\ 7&181.67&-1.88&22.47&\\
8&181.68&-2.39&14.61&\\ 9&181.76&-1.88&25.00&\\
10&181.78&-2.22&35.00&$\alpha_{4.85}^{1.42}=0.2$\\
11&181.81&-1.58&27.00&\\ 12&181.84&-1.56&18.22&\\
13&181.95&-2.35&11.80&\\ 14&182.02&-1.89&20.32&\\
15&182.05&-1.80&53.40&\\ 16&182.18&-2.01&37.30&\\
17&182.18&-1.40&33.00&\\ 18&182.27&-2.22&8.70&\\
19&182.27&-1.73&18.88&\\
20&182.32&-1.71&9.91&$\alpha_{1.42}^{0.151}=-2.27$\\
21&182.33&-2.35&10.30&\\ 22&182.35&-2.38&6.00&\\
23&182.38&-1.72&225.40&$\alpha_{0.151}^{0.365}=-0.86$\\
&&&&$\alpha_{1.42}^{0.408}=-0.75$\\ 24&182.40&-1.83&18.72&\\
25&182.41&-2.41&20.67&\\ 26&182.43&-2.44&11.90&\\
27&182.44&-2.53&22.33&\\
28&182.50&-2.48&57.80&$\alpha_{1.42}^{0.408}=-1.33$\\
29&182.58&-1.59&41.86&\\
\\
\hline
\end{tabular}
\end{table}

In Figure 2, lower panel, we show a 1.4-GHz VLA map made with data
from the NVSS Sky Survey (Condon et al. 1998), where all
point-like radio sources within the 68\% confidence contour of 3EG
J0542+2610 can be seen. The measured characteristics of these 29
sources are listed in Table 1. Most of them have no entry in any
existing point source catalog. In those cases where we were able
to find positional counterparts at other frequencies we have
estimated the spectral indices, which are also shown in the table.
Most of these sources are very weak, at the level of a few mJy. No
strong (at Jy level), flat or nonthermal source is within the
location error box of the gamma-ray source. The strongest radio
source (No. 23 in our table) has a rather steep spectrum and is a
factor $\sim 10$ below the minimum flux density of firm gamma-ray
blazar identifications given by Mattox et al. (1997). The nature
of this source is not clear at present; it could be a background
weak radio quasar. The fact that it is not seen at X-rays seems to
argue against a galactic microquasar or any other kind of
accreting source.

At X-ray energies the most significant source within the EGRET
error box is the X-ray transient A0535+26, which is discussed in
the next section. This source does not present significant radio
emission and consequently it cannot be seen in our maps. We have
indicated its position with a star symbol in Fig. 2, middle panel.
We also show in this figure the direction of the proper motion of
the system, as determined by Lee Clark \& Dolan (1999). The upper
panel shows the entire radio field as determined from radio
observations with Effelsberg 100-m single dish telescope at 1.408
GHz (data from Reich et al. 1997). The middle panel present an
enhanced image obtained at 2.695 GHz with the same telescope
(Fuerst et al. 1990), where the gamma-ray location probability
contours have been superposed (Hartman et al. 1999). We have
processed these large-scale images using the background filtering
techniques described by Combi et al. (1998).

\begin{figure}
\resizebox{9cm}{!}{\includegraphics{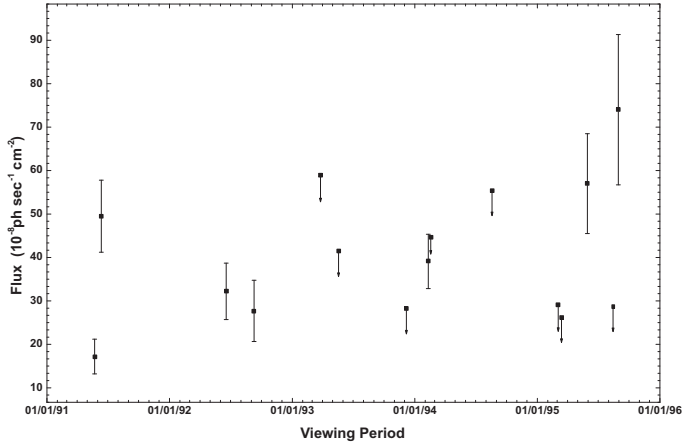}} \caption{\rm Flux
evolution of 3EG J0542+2610 through single viewing periods.}
\label{fig1}
\end{figure}

\begin{figure}
\caption{Radio maps at different resolutions of the field
containing the gamma-ray source 3EG J0542+2610. {\bf Upper panel}:
1408-MHz image obtained with Effelsberg 100-m single dish
telescope. The shell-type SNR G180.0-1.7 is clearly visible in the
map. Contours are shown in steps of 25 mJy beam$^{-1}$, starting
from 20 mJy beam$^{-1}$. {\bf Middle panel}: 2695-MHz map obtained
by the same telescope. Contours in steps of 30 mJy beam$^{-1}$,
starting from 30 mJy beam$^{-1}$. The position of A0535+26 is
marked by a star symbol. The arrow indicates the direction of the
proper motion. EGRET confidence location contours are superposed
to the radio image. {\bf Lower panel}: VLA image of the inner
region at 1.4 GHz. Contours in steps of 1 mJy beam$^{-1}$,
starting from 1 mJy beam$^{-1}$. The diffuse background emission
has been removed from the first two maps.} \label{fig2}
\end{figure}

\section{The X-ray transient A0535+26}

A0536+26 is a Be/X-ray transient where the compact object is a
104s pulsar in an eccentric orbit around the B0III star HDE 245770
(Giovanelli \& Sabau Graziati 1992). Be stars are rapidly rotating
objects which eject mass irregularly forming gaseous disks on
their equatorial planes. If there is a compact companion in a
close orbit, accretion from the star can result in strong X-ray
emission. In the case of A0535+26, strong and recurrent X-ray
outbursts are observed with a period of 111 days, which has been
identified with the orbital period (Giovanelli \& Sabau Graziati
1992). It is generally agreed that these outbursts occur when the
accretion onto the neutron star increases at the periastron
passage. The average ratio of the X-ray luminosity at the
periastron to that of the apoastron is $\sim100$ (Janot-Pacheco et
al. 1987). In Table 2 we list the main characteristics of the
A0535+26 system.

During a major outburst of A0535+26 in 1994, the BATSE instrument
of the Compton Gamma Ray Observatory detected a broad
quasi-periodic oscillation (QPO) in the power spectra of the X-ray
flux (Finger et al. 1996). The QPO component was detected during
33 days, with a central frequency that was well correlated with
both the hard X-ray flux and neutron star spin-up rate inferred
from pulse timing. Finger et al.'s (1996) observations are the
first clear evidence that an accretion disk is formed during giant
outbursts.

Using the simultaneous variations of the spin and the QPO
frequencies in the context of the beat frequency model (Alpar \&
Shaham 1985), Li (1997) has determined the evolution of the ratio
$\xi$ of the inner accretion disk radius to the Alfv\'en radius.
He found that during the initial rise of the outbursts $\xi$
quickly increased from $\sim 0.6$ to $\sim 1$, indicating a
transition of the accretion process from spherical accretion
before the outbursts to disk accretion during the high X-ray
luminosity phase.

Prior to the direct evidence for a transient accretion disk in
A0535+26, Motch et al. (1991) have already suggested, on the basis
of an analysis of the long-term X-ray, UV, and optical history of
the system, that two different types of interactions exist between
the Be star and the pulsar, accordingly with the dynamical state
of the highly variable circumstellar envelope. ``Normal" outbursts
would occur for high equatorial wind velocities ($\sim 200$ km
s$^{-1}$), whereas ``giant" outburst would result from lower
equatorial wind velocities ($\sim 20-80$ km s$^{-1}$) which allow
the formation of a transient accretion disk.

It is interesting to notice that, during the initial stage of the
1994 outburst, when QPOs were found in A0535+26, the gamma-ray
source 3EG J0542+2610 was detected by EGRET with a flux of
$(39.1\pm12.5)\;10^{-8}$ ph cm$^{-2}$ s$^{-1}$ (viewing period
321.1: February 8 - 15, 1994). But when A0535+26 was at the peak
of its X-ray luminosity, on Ferbuary 18, the gamma-ray source was
not detected (viewing period 321.5). The gamma-ray emission seems
to have been quenched precisely when the accretion disk was
well-formed and maximally rotating. In the next section we present
a model that can account for the gamma-ray production in A0535+26
necessary to explain the EGRET source 3EG J0542+2610 and that is
in agreement with our present knowledge of the Be/X-ray transient
source.

\begin{table}
\caption[]{Physical parameters for A0535+26 (from Janot-Pacheco et
al. 1987 and Giovanelli \& Sabau Graziati 1992)}
\begin{flushleft}
\begin{tabular}{l l}
\hline Primary spectral type & B0III \cr Primary mass & 9 - 17
$M_{\odot}$ \cr Secondary mass & $<2.7\;M_{\odot}$\cr Primary mass
loss rate & $7.7\;10^{-7}$ $M_{\odot}$ yr$^{-1}$ \cr  Distance &
$2.6\pm0.4$ kpc \cr $L_X$ peak (average) & $7.5\pm2.4\;10^{36}$
erg s$^{-1}$ \cr X-ray pulse period & 104 s \cr Orbital period &
$111\pm0.5$ d \cr Orbital eccentricity & 0.3 - 0.8 \cr $L_X^{\rm
max}/L_X^{\rm min}$ & $\sim 100$ \cr \hline
\end{tabular}
\end{flushleft}
\end{table}

\section{Model}

Our purpose in this section is to show that there exist a
plausible mechanism that could explain the gamma-ray emission of
the source 3EG J0542+2610 as originated in the X-ray transient
A0535+26. This mechanism should be capable of predicting the
observed gamma-ray flux, the variability in the lightcurve, the
fact that no gamma-ray emission was observed on February 18, when
A0535+26 was at the peak of the X-ray outburst, but also that it
was positively detected in the previous days when the X-ray flux
was rising, and finally the fact that A0535+26 is not a
non-thermal radio source. This latter restriction seems to suggest
a hadronic origin for the gamma rays. Otherwise, relativistic
electrons should also produce synchrotron radio emission.

Cheng \& Ruderman (1989, 1991) have studied the disturbances
produced in the magnetosphere of an accreting pulsar when the
Keplerian disk rotates more rapidly than the star. As in the case
of equal angular velocities, when $\Omega_{*}<\Omega_{\rm d }$,
inertial effects of electrons ($-$) and ions ($+$) lead to a
complete charge separation around the ``null surface" ${\bf
\Omega_{*}\cdot B}=0$. However, since now the equatorial plasma
between the inner accretion disk radius $r_0$ and the Alfv\'en
radius $r_{\rm A}$ co-rotates with the disk, whereas the rest of
plasma co-rotates with the star, an electrostatic gap with no
charge at all is created around the ``null surface" (see Figure
3). In this gap ${\bf E\cdot B}\neq 0$ and a strong potential drop
is established (see Cheng \& Ruderman 1991 for details).

\begin{figure}
\resizebox{9cm}{!}{\includegraphics{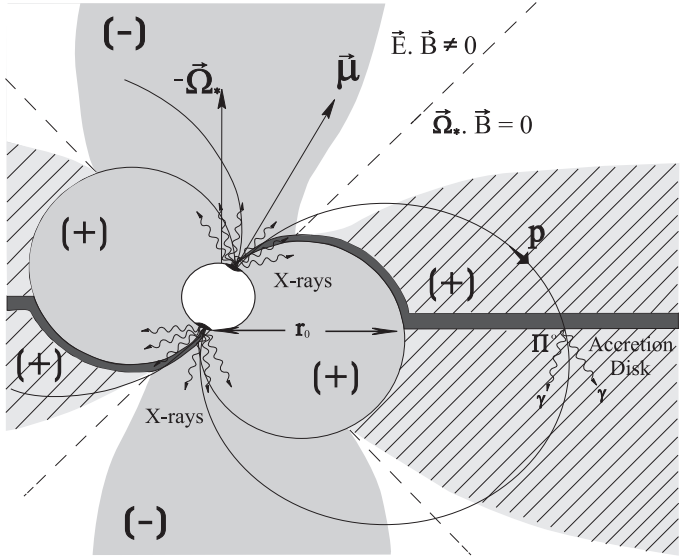}} \caption{Sketch of
the magnetosphere model for A0535+26 when $\Omega_{\rm
d}>\Omega_{*}$. The gray region co-rotates with the star, whereas
the hatched region co-rotates with the accretion disk. An empty
plasma gap where ${\bf E\cdot B}\neq 0$ exists around ${\bf
\Omega_{*}\cdot B}=0$. Protons from the stellar surface are
accelerated in this gap and collide with the disk, where they
produce $\pi^0$-decay gamma-rays. Adapted from Cheng \& Ruderman
(1991).} \label{fig3}
\end{figure}

Cheng \& Ruderman (1989) have shown that the potential drop along
the magnetic field lines through the gap is:
\begin{eqnarray}
\Delta V_{\rm max}&\sim&\frac{B_{\rm s}R^3\Omega_{\rm d}(r_0)}{r_0
c }\nonumber\\ &\sim&4\;10^{14}
\beta^{-5/2}\left(\frac{M}{M_{\odot}}\right)^{1/7}\!\!
R_6^{-4/7}L_{37}^{5/7}B_{12}^{-3/7}\;\;{\rm V},
\end{eqnarray}
where $B_{\rm s}$ is the neutron star's surface dipole magnetic
field, $R$ is its radius, $r_0$ is the inner accretion disk
radius, $M$ is the compact star mass, and $L_{37}$ is the X-ray
luminosity in units of $10^{37}$ erg s$^{-1}$. The radius and
magnetic field in the second expression are in units of $10^6$ cm
and $10^{12}$ Gauss, respectively. The parameter
$\beta\equiv2r_0/r_{\rm A}$ is twice the ratio of the inner
accretion disk radius to the Alfv\'en radius. For the particular
case of A0535+26 we can adopt $\beta\sim 1$, according to the
estimates by Li (1997).

Protons entering into the gap are accelerated up to energies above
$E_p\sim e\Delta V_{\rm max}$ whereas inverse Compton and
curvature losses would limit the energy gain of electrons and
positrons to lower values. The maximum current that can flow
through the gap can be determined from the requirement that the
azimuthal magnetic field induced by the current cannot exceed that
of the initial field ${\bf B}$ (Cheng \& Ruderman 1989):
\begin{eqnarray}
J_{\rm max}&\sim& cB_{\rm s}R^3 r_0^{-2}\nonumber\\
&\sim&1.5\;10^{24}
\beta^{-2}\left(\frac{M}{M_{\odot}}\right)^{-2/7}\!\!\!\!
R_6^{1/7}L_{37}^{4/7}B_{12}^{-1/7}\;{\rm esu\;s^{-1}}. \nonumber\\
\end{eqnarray}

As it can be seen from Fig. 3, the proton current flux is directed
from the polar cap of the star, where the accreting material
impacts producing strong X-ray emission and abundant ions to be
captured by the gap, towards the accretion disk. Electrons move in
the opposite sense in order to cancel any net charge current flow,
keeping in this way ${\bf E.B}\neq0$ in the gap.

The collision of the relativistic proton beam into the disk will
produce hadronic interactions with copious $\pi^0$ production
(Cheng et al. 1990). These $\pi^0$ will quickly decay into
$\gamma$-rays that could escape only if the disk density is
sufficiently low. Otherwise, they will be absorbed by the matter
and then re-emitted as X-rays. The interaction of relativistic
protons with a thin hydrogen layer has been studied by Cheng et
al. (1990) in the context of their model for the Crab pulsar. The
$\pi^0$-decay $\gamma$-rays will only escape insofar as the column
density of the disk would be $\Sigma\leq100$ g cm$^{-2}$. From
Novikov \& Thorne (1972), we have:
\begin{eqnarray}
\Sigma&\approx&7\;10^{2}
\beta^{-3/5}\left(\frac{M}{M_{\odot}}\right)^{-3/35}\!\!\!\!R_6^{-30/35}
L_{37}^{27/35}B_{12}^{-12/35}{\rm g\;cm^{-2}}.\label{mu}\nonumber
\\
\end{eqnarray}

In a system like A0535+26, the column density of the disk will
evolve with time following the variations in the X-ray luminosity.
All other physical parameters in Eq. (\ref{mu}) remain constant
along the orbital period. If at some point near the periastron,
when the X-ray luminosity is close to its maximum, the column
density exceeds the value $\sim 100$ g cm$^{-2}$, the medium will
be no longer transparent to gamma-rays and the gamma-emission will
be quenched (we are assuming that the disk is seen from the
opposite side to that where the current impacts). This could
explain the fact that A0535+26 has not been detected by EGRET at
the peak of its X-ray outbursts on February 18, 1994. The
detection, instead, was clear a few days before, during the
previous viewing period.

The expected hadronic $\gamma$-ray flux from A0535+26 on Earth
will be:

\begin{equation}
F(E_{\gamma})=\frac{1}{\Delta \Omega
D^2}\int^{\infty}_{E_{\gamma}}
\frac{d^{2}N_p(E_{\pi^0})}{dt\;dE_{\pi^0} }\;dE_p,\label{flux}
\end{equation}
where $D$ is the distance to the source, $\Delta \Omega$ is the
beaming solid angle and:
\be
\frac{d^{2}N_p(E_{\pi^0})}{dt\;dE_{\pi^0} }=\frac{2\,J_{\rm
max}}{e}\frac{\Sigma
<m_{\pi^0}>}{m_p}\frac{d\sigma(E_p,\,E_{\pi^0})}{dE_{\pi^0}} \ee
is the differential production rate of neutral pions from $pp$
interactions. In this latter expression,
$d\sigma(E_p,\,E_{\pi^0})/dE_{\pi^0}$ is the differential
cross-section for the production of $\pi^0$-mesons of energy
$E_{\pi^0}$ by a proton of energy $E_p$ in a $pp$ collision, and
$<m_{\pi^0}>$ is the mean multiplicity of $\pi^0$.

For calculation purposes we shall adopt the cross section given by
Dermer (1986a, b) and a beaming factor  $\Delta
\Omega/4\pi\approx0.3$ as used by Cheng et al. (1991). We shall
evaluate the total gamma-ray luminosity between 100 MeV and 20 GeV
of A0535+2610 at an early epoch $t$ of the disk formation, when
the X-ray luminosity ratio is $L_X^{\rm max}/L_X^{t}\sim 10$. If
the absorption feature observed near 110 keV is a cyclotron line,
the polar magnetic field of the neutron star results $9.5\;
10^{12}$ G (Finger et al. 1996). In our calculations we adopt this
value along with $\beta\sim 1$, $R_6=1$ and $M=1.4\;M_{\odot}$
(Janot-Pacheco et al. 1987). We then obtain that the column
density of the disk is $\sim42.5$ g cm$^{-2}$ at this stage and,
consequently, gamma-rays are not absorbed in the disk material.
The potential drop in the electrostatic gap results $\Delta V_{\rm
max}\sim2.5\;10^{13}$V whereas the proton current deposited into
the disk is $N_p=J_{\rm max}/e\sim4.8\;10^{32}$ s$^{-1}$. Using
Eq. (\ref{flux}) for photons in EGRET's energy range we obtain:

\begin{equation}
F(100\;{\rm MeV}<E_{\gamma}<20\;{\rm
GeV})\approx5.1\;10^{-8}\;\;{\rm ph\;cm^{-2}\;s^{-1}}.
\end{equation}
This flux is consistent with the lower EGRET detections. As the
accretion onto the neutron star increases the gamma-ray flux also
increases, reaching a maximum when $\Sigma\approx100$ g cm$^{-2}$.
At this point the flux has grown by a factor $\sim 4$ (notice the
dependence on $J_{\rm max}$ in addition to $\Sigma$) and photon
absorption into the disk becomes important quenching the
radiation. The gamma-ray source, then, is not detected when the
peak of the X-ray luminosity occurs. The additional gamma-ray flux
expected from secondary electrons and positrons radiating in the
the strong magnetic fields anchored in the disk can explain even
higher EGRET fluxes (Cheng et al. 1991). A more detailed analysis
in this sense will be presented elsewhere, including spectral
considerations. The observed spectrum depends on the proton
injection spectrum at the gap. Strong shocks near the polar gap
can accelerate the protons up to a power law which will be
preserved through the propagation across the gap and imprinted
into the $\pi^0$ gamma-ray spectrum. Additionally, $e^+e^-$
cascades induced in the disk material can result in a relativistic
bremsstrahlung contribution. Electromagnetic shower simulations
are in progress in order to determine this contribution (Romero et
al., in preparation).

\section{Discussion}

The model here outlined does not imply strict periodicity because
of the highly chaotic nature of the Be stellar winds, which can
significantly vary on short timescales producing strong changes in
the accretion rate. Notwithstanding, the general prediction that
the peak of the gamma-ray emission should not be coincident with
the maximum X-ray luminosity during a given outburst can be used
to test the general scenario proposed here. At present, the poor
time resolution of the gamma-ray lightcurve does not allow
correlation studies. EGRET data for the best sampled X-ray
outburst (in February 1994) consist only of two viewing periods,
as it was mentioned, and in one of them the source was not
detected. In the future, however, new instruments like GLAST could
provide the tools for these kind of investigations.

TeV emission should be produced in the accretion disk according to
Eq. (\ref{flux}), although degradation effects during the
propagation in the strong magnetic and photon fields around the
accretion disk could suppress much of it. The magnetic field at
the inner accretion disk radius is (e.g. Cheng et al. 1991):
\begin{equation}
B(r_0)=3\;10^5 \beta^{-3} B_{12}^{-5/7} R_6^{-9/7} L_{37}^{6/7}
(M/M_{\odot})^{-3/7}\;\;\;\;{\rm G} \label{B}
\end{equation}
which for the typical values of A0535+26 yields $B\sim10^4$ G. TeV
gamma-rays can be absorbed in the field through one-photon pair
production above the threshold given by (e.g. Bednarek 1993):
\begin{equation}
\chi\equiv \frac{E_{\gamma}B\sin{\theta}}{2m_e c^2 B_{\rm
cr}}=\frac{1}{15},
\end{equation}
where $\theta$ is the angle between the photon momentum and the
magnetic field and $ B_{\rm cr}$ is the critical magnetic field
given by  $B_{\rm cr}=m^2 c^3/ e\hbar\approx 4.4\;10^{13}$ G. This
process, then, suppress gamma-ray photons with energies higher
than $\sim300$ TeV in A0535+26. Since this value is well above the
energy of the protons that impact on the disk, we find that
absorption in the magnetic field should not occur.

However, TeV photons with lower energies should be absorbed by
two-photon pair production in the accretion disk X-ray
photosphere. Calculations by Bednarek (1993) show that the optical
depth quickly goes to values above 1 for disks with luminosities
$L_{37}\sim 1$. The opacity effects can reach even GeV energies
leading to a steepening in the spectrum respect to what is
expected from a pure pion-decay mechanism. The fact that the
observed spectrum in 3EG J0542+4610 has an index $\Gamma\sim -2.7$
seems to support the idea that important absorption is occurring
in the X-ray photosphere of this source. Additional spectral
modifications should be produced by the emission of secondary
pairs in the magnetic field close to the disk (Cheng et al. 1991,
Romero et al., in preparation).

In their original model, Cheng \& Ruderman (1989) suggested that
if the disk is sufficiently dense then the gamma-rays could be
produced only in a moving low-density ``window". This window would
collimate a pencil beam of gamma-rays aligned with the magnetic
axis. Consequently, a pulsed emission could exist with the same
period of the X-ray source (104 sec in the case of A0235+26).
Small changes in the gamma-ray period might be produced by the
radial motion of the ``window". It would be interesting to test
whether this pulses are present in A0235+26. However, since the
disk is a transient structure and only would excess the critical
density during a few days per orbit, the number of photon counts
in EGRET data are too low to allow a periodicity analysis as in
the case of isolated and stable gamma-ray pulsars. Instruments
with higher sensitivity like GLAST could sum up over several
viewing periods in order to look for these features.
\\

\section{Summary}

We have shown that the Be/X-ray transient system A0535+26 can be
also a transient gamma-ray source under very reasonable
assumptions. The existence of a QPO phenomenon detected by BATSE
during the 1994 X-ray outburst provided direct evidence of the
formation of a transient accretion disk near the periastron
passage. In the beat frequency model for QPO the Keplerian orbital
frequency of the material at the inner edge of the accretion disk
should exceed the spin frequency of the neutron star. Cheng \&
Ruderman (1989) have shown that in such a circumstance an
electrostatic gap is open in the magnetosphere around the null
surface determined by ${\bf \Omega_*\cdot B=0}$. This gap can
accelerate protons up to energies of tens of TeV, producing a
hadronic current that impacts into the accretion disk generating
$\pi^0$ gamma-rays. The transient character of the disk makes the
high-energy gamma radiation highly variable, as was found in the
observed gamma-ray flux evolution. A specific prediction of the
model is the suppression of gamma-ray emission when the column
density of the disk exceeds a critical value, near the peak of the
X-ray luminosity. Future GeV and TeV observations of this source
with instruments of high temporal resolution, like GLAST or 5@5
(see Aharonian et al. 2001), could be used to test the proposed
model and, if it is basically correct, to probe the evolution of
the matter content on the accretion disk in this extraordinary
X-ray binary.

\begin{acknowledgements}
GER thanks valuable comments by Dr. Pablo H. Posta. An anonymous
referee made interesting suggestions that lead to a substantial
improvement of the manuscript. This work was supported by CONICET
(PIP 0430/98), ANPCT (PICT 03-04881) and Fundaci\'on Antorchas.
\end{acknowledgements}

{}

\end{document}